\documentstyle[epsf,aps]{revtex}
\begin{document}

\twocolumn[\hsize\textwidth\columnwidth\hsize\csname 
@twocolumnfalse\endcsname

\title{Nagel Scaling,  Relaxation and Universality in the Kinetic Ising Model on an
Alternating Isotopic Chain}
\author{L. L. Gon\c {c}alves}
\address{Departamento de Fisica,
Universidade Federal do Cear\'{a},
Fortaleza, Cear\'{a}, Brazil}
\author{M. L\'{o}pez de Haro\cite{mariano}}
\address{Centro de Investigaci\'on en Energ\'{\i}a, UNAM, 
Temixco, Morelos 62580, M\'{e}xico}
\author{J. Tag\"{u}e\~{n}a-Mart\'{\i}nez\cite{julia}}
\address{Facultad de
Ciencias,
Universidad Aut\'onoma del Estado de Morelos, 
Cuernavaca,
Morelos 62210, M\'{e}xico}
\author{R. B. Stinchcombe}
\address{Department
of Theoretical Physics,
Oxford University,
1 Keble Road, Oxford OX1 3NP,
United Kingdom}
\date{14 June 1999}
\maketitle

\begin{abstract}
The
dynamic critical exponent and the frequency and wave-vector
dependent
susceptibility of the kinetic Ising model on an alternating
isotopic chain\
with Glauber dynamics are examined. The analysis provides
to our knowledge
the first connection between a microscopic model and the
Nagel scaling curve
originally proposed to describe dielectric
susceptibility measurements of
several glass-forming liquids. While support
is given to the hypothesis
relating the Nagel scaling to multiple
relaxation processes, it is also found that
the scaling function may exhibit plateau
regions and does not hold for all
temperatures.
\end{abstract}

\vskip2pc] \narrowtext

Experimental work on dielectric relaxation in glass-forming liquids has in
recent years been reported in terms of a new (thought to be universal)
scaling function \cite{Nagel} which is presumed to be related to
multifractal scaling. While the more usual normalized Debye scaling in terms
of a single relaxation time is very simple (one chooses to scale the
frequency with the inverse of the relaxation time and the real and imaginary
parts are then divided by their values at zero and one, respectively), in
the so-called Nagel plot the abscissa is $(1+W) \log _{10}\left(
\omega/\omega _p\right)/W^2 $ and the ordinate is $\log _{10}\left( \chi
^{\prime \prime }(\omega )\omega _{p}/\omega \Delta \chi \right)/W $. Here, $%
\chi ^{\prime \prime }$ is the imaginary part of the susceptibility, $W$ is
the full width at half maximum of $\chi ^{\prime \prime }$, $\omega $ is the
frequency and $\omega _{p}$ the one corresponding to the peak in $\chi
^{\prime \prime }$, and $\Delta \chi =\chi (0)-\chi _{\infty }$ is the
static susceptibility. Despite its undeniable phenomenological success, such
scaling is not quite well understood on a physical basis. The authors of
this proposal advance the idea that the presence of more than one relaxation
process is not alien to this form and thus suggest that multifractality such
as the one present in theories of chaos may well be behind the new scaling.
In order to gain some insight into the physical origin of the Nagel scaling,
it seems appropriate to consider simple but well established models in which
both universal features are unquestionable and more than one relaxation
mechanism is present. A good candidate may be found among kinetic Ising
models.

The scaling hypothesis of Halperin and Hohenberg \cite{Hohenberg} relates
the time scale $\tau $ and the correlation length $\xi $ and introduces the
dynamic critical exponent $z$. In the case of Ising models, $z$ was for a
long time believed to be universal, depending on the nature of conserved
quantities and of those features, for instance dimensionality \cite{Zhu},
which determine their static universality class. However, it is now well
established that for some simple systems this exponent is non universal\cite
{Kimball}-\cite{Robin}. In particular, in the case of one-dimensional
Glauber dynamics \cite{Glauber} the alternating isotopic chain\cite{Lind1}
presents universal behavior (in the sense that it leads to the same value of
the dynamic critical exponent as the homogeneous chain) whereas the
alternating-bond chain does not \cite{Lind1}-\cite{Tong} (see however Ref. 
\cite{Southern}). Due to the fact of this universality of $z$, the isotopic
alternating Ising chain with Glauber dynamics provides a test model in which
to assess the value of the connection between multiple relaxation mechanisms
and the Nagel plot. The model consists of a closed linear chain with N sites
occupied by two isotopes (characterized by two different spin relaxation
times) that are alternately arranged. The Hamiltonian is the usual Ising
Hamiltonian given by

\begin{equation}
H=-J\sum\limits_{j=1}^{N}\sigma _{j}\sigma _{j+1},  \label{1}
\end{equation}

\noindent where $\sigma _{j}$ is a stochastic (time-dependent) spin variable
assuming the values $\pm 1$ and $J$ the coupling constant. The configuration
of the chain is specified by the set of values $\left\{ \sigma _{1},\sigma
_{2},...\sigma _{N}\right\} \equiv \left\{ \sigma ^{N}\right\} $ at time $t$%
. This configuration evolves in time due to interactions with a heat bath.
We assume for this chain the usual Glauber dynamics so that the transition
probabilities are given by

\begin{equation}
w_{i}(\sigma _{i})=\alpha _{i}\left( 1-\frac{\gamma }{2}(\sigma _{i-1}\sigma
_{i}+\sigma _{i}\sigma _{i+1})\right) ,  \label{3b}
\end{equation}

\noindent where $\gamma =\tanh \left( 2J/k_BT \right) $, $k_{B}$ being the
Boltzmann constant and $T$ the absolute temperature, and $\alpha _{i}$ is
the inverse of the relaxation time $\tau _{i}$ of spin $i$ in the absence of
spin interactions.

If we now let $\alpha _{1}$ and $\alpha _{2}$ represent the inverses of the
free spin relaxation times of chains composed solely of spins of species $1$
or species $2$, respectively, then we can set $\alpha _{i}=\overline{\alpha }%
_{1}-(-1)^{i}\overline{\alpha _{2}}$ , where $\overline{\alpha }_{1}=(
\alpha _{1}+\alpha _{2})/2$ and $\overline{\alpha }_{2}=(\alpha _{1}-\alpha
_{2})/2$.

The time dependent probability $P\left( \left\{ \sigma ^{N}\right\}
,t\right) $ for a given spin configuration satisfies the master equation

\begin{eqnarray}
\frac{dP\left( \left\{ \sigma ^{N}\right\} ,t\right) }{dt}
&=&-\sum\limits_{i=1}^{N}w_{i}\left( \sigma _{i}\right) P\left( \left\{
\sigma ^{N}\right\} ,t\right)  \nonumber \\
&&+\sum\limits_{i=1}^{N}w_{i}\left( -\sigma _{i}\right) P\left( T_{i}\left\{
\sigma ^{N}\right\} ,t\right) ,  \label{4}
\end{eqnarray}
where $T_{i}\left\{ \sigma ^{N}\right\} \equiv \{\sigma _{1},\sigma
_{2},...\sigma _{i-1},-\sigma _{i},\sigma _{i+1},...\sigma _{N}\}$. The
dynamical properties we are interested in, namely the dynamic critical
exponent and the susceptibility, require the knowledge of some moments of
the probability $P\left( \left\{ \sigma ^{N}\right\} ,t\right) $. Hence, we
introduce the following expectation values and correlation functions defined
as:

\begin{equation}
q_i\left( t\right) =\left\langle \sigma _i\left( t\right) \right\rangle
=\sum\limits_{\left\{ \sigma ^N\right\} }\sigma _iP\left( \left\{ \sigma
^N\right\} ,t\right) ,  \label{5}
\end{equation}

\begin{equation}
r_{i,j}\left( t\right) =\left\langle \sigma _i\left( t\right) \sigma
_j(t)\right\rangle =\sum\limits_{\left\{ \sigma ^N\right\} }\sigma _i\sigma
_jP\left( \left\{ \sigma ^N\right\} ,t\right) ,  \label{6}
\end{equation}

\noindent and

\begin{eqnarray}
&&c_{i,j}\left( t^{\prime },t^{\prime }+t\right) =\Theta \left( t\right)
\left\langle \sigma _{i}\left( t^{\prime }\right) \sigma _{j}\left(
t^{\prime }+t\right) \right\rangle  \nonumber \\
&=&\sum\limits_{\left\{ \sigma ^{N}\right\} ,\left\{ \sigma ^{N\prime
}\right\} }\sigma _{i}^{\prime }P\left( \left\{ \sigma ^{N\prime }\right\}
,t^{\prime }\right) \sigma _{j}p\left( \left\{ \sigma ^{N}\right\} |\left\{
\sigma ^{N\prime }\right\} ,t\right) ,  \label{7}
\end{eqnarray}

\noindent where $\Theta \left( t\right) $ is the Heaviside step function and
the sums run over all possible configurations. The second equality of Eq. (%
\ref{7}), which gives the formal definition of the time-delayed correlation
function, involves $p\left( \left\{ \sigma ^{N}\right\} |\left\{ \sigma
^{N\prime }\right\} ,t\right) $, the conditional probability of the chain
having the configuration $\left\{ \sigma ^{N}\right\} $ at time $t^{\prime
}+t$ provided it had the configuration $\left\{ \sigma ^{N\prime }\right\}
=\left\{ \sigma _{1}^{\prime },\sigma _{2}^{\prime },...,\sigma _{N}^{\prime
}\right\} $ at time $t^{\prime }$. Multiplying the master equation by the
appropriate quantities and performing the required summations we obtain the
set of time evolution equations that will be used in our later development.
These are given by

\begin{equation}
\frac{dq_{j}}{dt}=-\alpha _{j}\left( q_{j}-\frac{\gamma }{2}%
(q_{j-1}+q_{j-1})\right)  \label{8}
\end{equation}

\noindent and

\begin{eqnarray}
&&\frac{dc_{i,j}\left( t^{\prime },t^{\prime }+t\right) }{dt}%
=r_{i,j}(t^{\prime })\delta (t)-\alpha _{j}c_{i,j}\left( t^{\prime
},t^{\prime }+t\right)  \nonumber \\
&&+\frac{\alpha _{j}\gamma }{2}(c_{i,j-1}\left( t^{\prime },t^{\prime
}+t\right) +c_{i,j+1}\left( t^{\prime },t^{\prime }+t\right) ).
\end{eqnarray}

We now impose translational invariance and introduce $\widetilde{q}_k$, the
(spatial) Fourier transform of $q_j$, the $t^{\prime }\rightarrow \infty $
limit of the (temporal) Fourier transform of $c_l\left( t^{\prime
},t^{\prime }+t\right) \equiv c_{i,j}\left( t^{\prime },t^{\prime }+t\right) 
$ (with $l=j-i$) denoted by $\widehat{c}_l(\omega )$, and $\widetilde{C}%
_k(\omega )$, the spatial Fourier transform of $\widehat{c}_l(\omega )$,
defined through

\begin{equation}
q_j=\frac 1{\sqrt{N}}\sum\limits_k\widetilde{q}_k\exp \left( ikj\right) ,
\label{10a}
\end{equation}

\begin{equation}
\widehat{c}_l(\omega )=\lim_{t^{\prime }\rightarrow \infty }\frac 1{2\pi }%
\int_{-\infty }^\infty c_l\left( t^{\prime },t^{\prime }+t\right) \exp
(-i\omega t)dt  \label{11}
\end{equation}

\noindent and

\begin{equation}
\widetilde{C}_k(\omega )\equiv \left\langle \sigma _{-k}\sigma
_k\right\rangle _\omega =\frac 1N\sum_l\widehat{c}_l(\omega )\exp (-ikl).
\label{12}
\end{equation}

In terms of these quantities, Eqs. (\ref{8}) and (8) may be rewritten,
respectively, as

\begin{equation}
\frac{d\Psi _k}{dt}={\bf M}_k\Psi _k  \label{13}
\end{equation}
\noindent and

\begin{equation}
i\omega \widehat{c}_{l}(\omega )=r_{l}^{\infty }-\alpha _{l}\left( \widehat{c%
}_{l}(\omega )-\frac{\gamma }{2}(\widehat{c}_{l-1}(\omega )+\widehat{c}%
_{l+1}(\omega ))\right) ,  \label{14}
\end{equation}

\noindent where

\begin{equation}
\Psi _k = \left( 
\begin{array}{c}
\widetilde{q}_k \\ 
\widetilde{q}_{k-\pi }
\end{array}
\right) ,  \label{15}
\end{equation}

\begin{equation}
{\bf M}_{k}=\left( 
\begin{array}{cc}
-\overline{\alpha }_{1}\left( 1-\gamma \cos k\right) & \; -\overline{\alpha }%
_{2}\left( 1+\gamma \cos k\right) \\ 
-\overline{\alpha }_{2}\left( 1-\gamma \cos k\right) & \; -\overline{\alpha }%
_{1}\left( 1+\gamma \cos k\right)
\end{array}
\right) ,  \label{16}
\end{equation}

\noindent and $r_{l}^{\infty }={\lim }_{t\rightarrow \infty }r_{l}(t)$ is
the value of the pair correlation function corresponding to the stationary
solution of the equations of motion in the limit $t\rightarrow \infty $.

The solution to Eq. (\ref{13}), which yields the magnetization, is
straightforward, namely

\begin{equation}
\Psi _k\left( t\right) =e^{{\bf M}_kt}\Psi _k\left( 0\right) .  \label{17}
\end{equation}

The relaxation process of the wave-vector dependent magnetization is
determined by the eigenvalues of ${\bf M}_k$ . These are given by

\begin{equation}
\lambda _{k}^{\pm }=-\overline{\alpha }_{1}\pm \sqrt{\overline{\alpha }%
_{2}^{2}+\left( \overline{\alpha }_{1}^{2}-\overline{\alpha }_{2}^{2}\right)
\gamma ^{2}\cos ^{2}k}.  \label{18}
\end{equation}

The inverses of the ($k$-dependent ) relaxation times $\tau _{k}^{\pm }$ of
the $\pm k^{th}$ modes are precisely the $\lambda _{k}^{\pm }$. In the
critical region, that is when $T\rightarrow 0$ and $k\rightarrow 0$, $%
\lambda _{k}^{-}\rightarrow -2\overline{\alpha }_{1}$ while $\lambda
_{k}^{+}\rightarrow 0.$ This means that the critical mode is the one
corresponding to $\lambda _{k}^{+}$. As for the relaxation time, in this
limit one gets

\begin{equation}
{\rm Re}\left( -\lambda _{k}^{+}\right) =-\frac{1}{\tau _{k}}\sim \frac{%
\left( \overline{\alpha }_{1}^{2}-\overline{\alpha }_{2}^{2}\right) }{2%
\overline{\alpha }_{1}}\xi ^{-2}\left[ 1+(\xi k)^{2}\right] ,  \label{19}
\end{equation}

\noindent where we have identified the correlation length $\xi $ as $\xi
=\exp (2J/k_{B}T)/2$. By comparing the former expression with the one of the
dynamic scaling hypothesis $\tau _{k}^{-1}\sim \xi ^{-z}f(\xi k)$, one finds 
$z=2$, so that, as stated above, in the case of one-dimensional Glauber
dynamics the alternating isotopic chain\cite{Lind1} leads to the same value
of $z$ as the homogeneous chain.

Now we turn to the calculation of the other interesting response function,
namely the frequency and wave-vector dependent susceptibility $S_{k}\left(
\omega \right) $, which, by virtue of the fluctuation-dissipation theorem 
\cite{Kubo}, is defined by 
\begin{equation}
S_{k}\left( \omega \right) =\frac{\left\langle \sigma _{k}\sigma
_{-k}\right\rangle _{\infty }}{k_{B}T}-\frac{i\omega \left\langle \sigma
_{k}\sigma _{-k}\right\rangle _{\omega }}{k_{B}T},  \label{20}
\end{equation}
where $\left\langle \sigma _{k}\sigma _{-k}\right\rangle _{\infty }=1/\left(
1-\gamma \cos k\right) \cosh (2J/k_{B}T)$ is the static correlation function
and $\left\langle \sigma _{k}\sigma _{-k}\right\rangle _{\omega }$ is the
Fourier transform of the dynamic one. \noindent It should be noted that $%
\chi \equiv k_{B}TS_{0}(\omega )/\left\langle \sigma _0\sigma
_0\right\rangle _{\infty }$. After some rather lengthy but not too
complicated algebraic manipulations starting with Eq.(\ref{14}) one may
arrive at the following result \cite{Lind1}, namely

\begin{eqnarray}
&&S_{k}\left( \omega \right) =\frac{1}{k_{B}T\left( 1-\gamma \cos k\right)
\cosh \frac{2J}{k_{B}T}}  \nonumber \\
&&\times \left[ 1-\frac{i\omega \left( i\omega +\overline{\alpha }%
_{1}(1+\gamma \cos k)\right) }{\left( i\omega +\overline{\alpha }_{1}\right)
^{2}-\frac{1}{2}\gamma ^{2}\alpha _{1}\alpha _{2}\left( 1+\cos 2k\right) -%
\overline{\alpha }_{2}^{2}}\right] ,  \label{21}
\end{eqnarray}
from which, using also Eq. (\ref{18}) with $k=0$, $\chi $ can be expressed
in the form

\begin{equation}
\chi =\frac{(1-\gamma )(\alpha _{1}+\alpha _{2})}{4}\left[ \frac{1-f(\alpha
_{1},\alpha _{2},\gamma )}{i\omega -\lambda _{0}^{+}}-\frac{1+f(\alpha
_{1},\alpha _{2},\gamma )}{i\omega -\lambda _{0}^{-}}\right] .  \label{22}
\end{equation}
Here, the (temperature dependent) function $f(\alpha _{1},\alpha _{2},\gamma
)$ is given by

\begin{equation}
f(\alpha _{1},\alpha _{2},\gamma )=\frac{(\alpha _{1}-\alpha
_{2})^{2}-4\alpha _{1}\alpha _{2}\gamma ^{2}}{(\alpha _{1}+\alpha _{2})\sqrt{%
(\alpha _{1}-\alpha _{2})^{2}+4\alpha _{1}\alpha _{2}\gamma ^{2}}}.
\label{23}
\end{equation}
If we set $\alpha _{1}=\alpha _{2}$ in Eq. (\ref{22}), {\it i.e. }we take
the uniform chain, then of course the resulting susceptibility has the
simple Debye form. Although not shown, we have checked that this form does
not lead to Nagel scaling. On the other hand, for the case $\gamma =0,$ we
get

\begin{equation}
\chi _{\gamma =0}=\frac{1}{2}\left[ \frac{\alpha _{1}}{i\omega +\alpha _{1}}+%
\frac{\alpha _{2}}{i\omega +\alpha _{2}}\right] ,  \label{24}
\end{equation}
so that the general structure of the result for the susceptibility of the
alternating isotopic chain is preserved irrespective of the value of $\gamma 
$ ({\it i.e.} of the temperature), namely a linear combination of two
Debye-like terms.

In Figs. 1 to 3 we present Nagel plots for the cases $\alpha _{1}=1$ and $%
\alpha _{2}=2$, $\alpha _{1}=1$ and $\alpha _{2}=10$, and $\alpha _{1}=1$
and $\alpha _{2}=1000$, respectively, and different values of $1/T^{\ast
}\equiv 2J/k_{B}T$. In the inserts we include the plots $\chi ^{\prime
\prime }(\omega )/\chi ^{\prime \prime }(\omega _{p})$ against $\omega
/\omega _{p}$ which are the natural variables of the Debye scaling. We note
that as soon as the relaxation times become different, except for very low
values of $T^{\ast }$, the agreement with the \ Nagel scaling improves
significantly ({\it cf. }Fig. 1) until such scaling is virtually perfect as
depicted in Figs. 2 and 3. On the other hand, the almost perfect Debye
scaling in Fig. 1 is completely lost in Fig.3. It should be noted that if
the two time scales are very different, plateau regions eventually appear in
the Nagel plot as clearly seen in Fig. 3. Whether the presence of more than
two relaxation times, even if not as widely separated as in the case of Fig.
3, would also lead to the same type of results is something requiring future
assessment. Also, the precise location of the critical value of $T^{\ast }$
above which the Nagel scaling holds as well as the nature of the crossover
and of the \ `low $T^{\ast }$' \ regime are worth investigating.

It is important to point out that the experiments in which the Nagel plots
have been more successful concern glass-forming systems in which topological
constraints are assumed to be crucial. However, a clearcut connection
between such constraints and the different relaxation mechanisms has not
been established. In this sense, it is rewarding that the alternating
isotopic Ising chain, which is relatively simple with regards to relaxation
phenomena but shows nevertheless universal behavior in terms of the dynamic
critical exponent, provides perhaps the {\it first microscopic model} in
which this scaling is shown to arise. We further want to mention that in
order to include some of the features present in systems such as the above
mentioned glass-forming liquids, we have also considered in the present
context a generalization to an alternating isotopic chain of our quasi
one-dimensional kinetic Ising-like model of linear polymeric chains \cite
{LHTGE}, in which the Hamiltonian was chosen as to reduce to the one giving
the intramolecular energy of the Gibbs-di Marzio lattice model \cite{GdM}.
Interestingly enough in this model, in which the stochastic dynamics implied
a rule of transition for the configurational changes which was tied to the
creation or disappearance of flexes and so only some states were selected
(in the magnetic language this means that the domain wall motion is through
a biased random walk), similar conclusions concerning the Nagel scaling
readily follow. These will be reported elsewhere. Finally, it would be
interesting to test whether the same kind of scaling is present in other
Ising models related to glassy systems, such as the spin facilitated kinetic
Ising model originally introduced by Fredrickson and Andersen \cite{FA} and
recently studied in connection with glassy dynamics \cite{Schulz}.

One of us (L. L. Gon\c{c}alves) wants to thank the Brazilian agencies CNPq
and FINEP for partial financial support. The work has also received partial
support by DGAPA-UNAM under projects IN103797 and IN104598.

\strut

{\bf Figure captions}

%
%\begin{figure}[htb]
%\begin{center}
%\epsfxsize=3.2in             %so many inches wide
%\leavevmode\epsfbox{Fig1.ps}
%\end{center}
%\caption{
%Nagel plot for $\alpha _{1}=1$ and $\alpha _{2}=2$ and for $%
%T^{\ast }=1,2,5,10$ and $100$. There is reasonable agreement with the
%scaling form for this choice except for low $T^{\ast }$.
%}%\label{}
%\end{figure}
%
%\begin{figure}[htb]
%\begin{center}
%\epsfxsize=3.2in             %so many inches wide
%\leavevmode\epsfbox{Fig2.ps}
%\end{center}
%\caption{
%The same as Fig. 1 but with the choice $\alpha _{1}=1$ , $\alpha
%_{2}=10$ and $T^{\ast }=5,10,50$ and $100$. The improvement in the agreement
%with the Nagel scaling is rather noticeable.
%}%\label{}
%\end{figure}
%
%\begin{figure}[htb]
%\begin{center}
%\epsfxsize=3.2in             %so many inches wide
%\leavevmode\epsfbox{Fig3.ps}
%\end{center}
%\caption{
%The same as Figs. 1 and 2 but for $\alpha _{1}=1$ , $\alpha
%_{2}=1000$ and $T^{\ast }=5,10,50$ and $100$. A plateau region is clearly
%present in this case.
%}%\label{}
%\end{figure}
%

Figure 1. Nagel plot for $\alpha _{1}=1$ and $\alpha _{2}=2$ and for $%
T^{\ast }=1,2,5,10$ and $100$. There is reasonable agreement with the
scaling form for this choice except for low $T^{\ast }$. The insert contains
the plot of $\chi ^{\prime \prime }(\omega )/\chi ^{\prime \prime }(\omega
_{p})$ vs. $\omega /\omega _{p}$ in order to test the Debye-like behavior.

Figure 2. The same as Fig. 1 but with the choice $\alpha _{1}=1$ , $\alpha
_{2}=10$ and $T^{\ast }=5,10,50$ and $100$. The improvement in the agreement
with the Nagel scaling is rather noticeable, while the opposite trend is
observed with respect to the Debye scaling.

Figure 3. The same as Figs. 1 and 2 but for $\alpha _{1}=1$ , $\alpha
_{2}=1000$ and $T^{\ast }=5,10,50$ and $100$. A plateau region in the Nagel
plot is clearly present in this case. Here, the behavior is definitely
non-Debye.

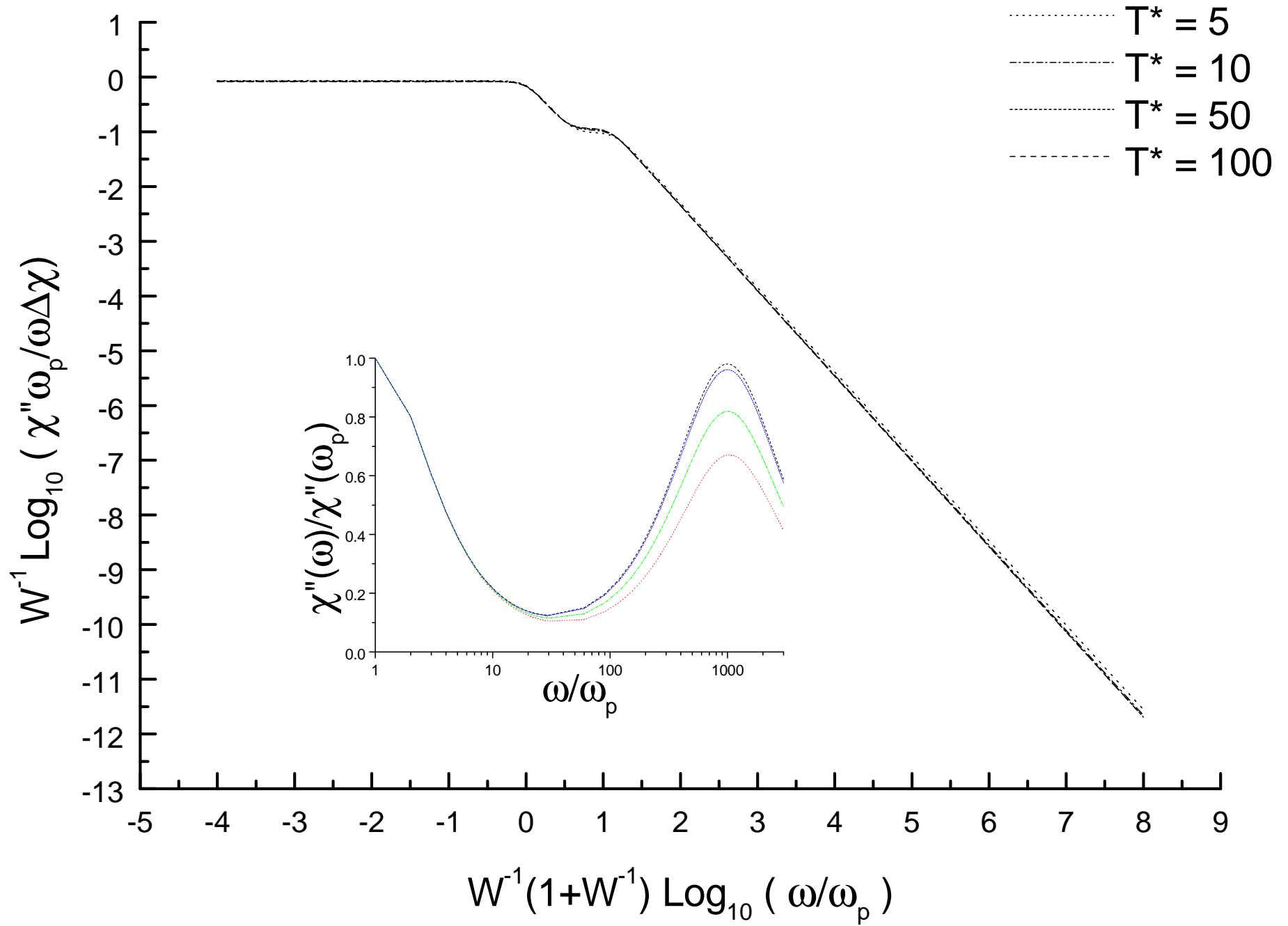
\end{document}